\documentclass[12pt]{article}

\usepackage{verbatim}
\usepackage{amsmath}
\usepackage{amssymb}
\usepackage{graphicx}
\usepackage{cite}
\usepackage{subfig}
\usepackage{setspace}
\usepackage{url}
\usepackage[percent]{overpic}
\usepackage{slashed}
\usepackage{authblk}
\usepackage{hyperref}

\usepackage{xspace}

\usepackage{fullpage}
\usepackage{hyperref}

\def\ie{{\it i.e.}}
\def\eg{{\it e.g.}}

\def\to{\rightarrow}

\title{Mass Reach Scaling for Future Hadron Colliders}
\date{}
\author{Thomas G. Rizzo}
\affil{SLAC National Accelerator Laboratory, Menlo Park, CA, 94025, USA\footnote{rizzo@slac.stanford.edu}}

\begin{document}

\rightline{\vbox{\halign{&#\hfil\cr
&SLAC-PUB-16206\cr
}}}


{\let\newpage\relax\maketitle}

\begin{abstract}
The primary goal of any future hadron collider is to discover new physics (NP) associated with a high mass scale, $M$, 
beyond the range of the LHC. In order to maintain the same {\it relative} mass reach for NP, $M/{\sqrt s}$, as $\sqrt s$ 
increases, Richter recently reminded us that the required integrated luminosity obtainable at future hadron colliders 
(FHC) must grow rapidly, $\sim s$, in the limit of naive scaling. This would imply, \eg, a $\sim 50$-fold increase in 
the required integrated luminosity when going from the 14 TeV LHC to a FHC with $\sqrt s=100$ TeV, an increase that would 
prove quite challenging on many different fronts. In this paper we point out, due to the scaling violations associated 
with the evolution of the parton density functions (PDFs) and the running of the strong coupling, $\alpha_s$, that the 
actual luminosity necessary in order to maintain any fixed value of the relative mass reach is somewhat greater than this 
scaling result indicates. However, the actual values of the required luminosity scaling are found to be dependent upon the 
detailed nature of the NP being considered. Here we elucidate this point explicitly by employing several specific 
benchmark examples of possible NP scenarios and briefly discuss the (relatively weak) search impact in each case if these 
luminosity goals are not met.  
\end{abstract}

\section{Introduction and Background}

There is rising interest in the physics potential of a future higher energy hadron collider which might begin running 
sometime after the high luminosity LHC program is completed. Such a machine, here generically termed the FHC, has been 
discussed in various manifestations at a growing number of workshops{\cite {FHC}} but generally is expected to operate 
in the $\sqrt s \sim 100$ TeV energy range. The goal of 
such a machine will be to explore the unknown, \ie, to search for new physics (NP), both beyond the Standard Model (SM) 
and beyond the reach of the LHC, that might be kinematically accessible at these higher collision energies. As such, 
it's NP search capabilities should be as strong as possible, and in particular be at least as powerful as those we 
expect to be available at the 14 TeV LHC. This NP may take several forms: it may be relatively light but very weakly 
coupled to the SM so that the the increased cross sections at a higher energy collider will allow access. More commonly, 
we imagine this NP to manifest as some new, very heavy state(s) that are simply too massive to be produced at the 14 
TeV LHC; this is the case we will consider below.  

Searches for NP can be quite complex, generally involving sophisticated experimental analyses in order to extract a 
significant signal above some SM background. This makes quantifying the power of a future collider difficult without 
a detailed study of a wide range of potential NP physics scenarios. Depending on what kind of NP one is interested 
in various possibilities come to mind. Here, as said above, we are essentially only interested in NP which is quite 
heavy. Perhaps in this case a crude but simple measure of this potential discovery power is obtainable by employing 
the value of the {\it relative} mass reach for NP associated with a heavy mass scale $M$, \ie,  the value of the mass 
reach scaled to the collision energy, $M/{\sqrt s}$.  For example, this would mean that if a new 3.5 TeV state is 
discoverable with some fixed number of signal events at the 14 TeV LHC then the corresponding 25 TeV state should 
be discoverable at the 100 TeV FHC with the same number of events.  We can achieve this by requiring that as the  
$\sqrt s$ increases the number of NP events remains at (or above) the given fixed reference value. To accomplish 
this it is clear that the integrated luminosity, $L$, of the collider must grow with increasing $\sqrt s$. Richter 
has recently emphasized this issue{\cite {Richter}} and reminded us that in the scaling limit for the NP cross section, 
$\sigma$, $L$ must grow as $\sim s$, the square of the collider center of mass energy.  In such a limit, the relevant 
ratio of appropriately {\it energy-scaled} cross sections is given by:         
\begin{equation}
R={{\sigma(M/{\sqrt s}, ~s)}\over {\sigma(M/{\sqrt s}, ~s=s_0)}} ~{{s}\over s_0}  \,
\end{equation}
which is defined for a {\it fixed} value of the ratio $M/{\sqrt s}$ and by a reference collision energy, ${\sqrt {s_0}}$. 
{\footnote {Here, since we will be making comparison with the LHC it will be convenient to take ${\sqrt {s_0}}=14$ TeV.}}  
Note that the value of $R$ is then simply unity in the scaling limit, by construction, reflecting the required 
luminosity growth as discussed above. 

Of course we know that exact scaling is broken in the real world arising from a 
number of sources, \eg,  due to the evolution of the parton density functions (PDFs) and the running of the strong coupling,  
$\alpha_s$.  (We note that the running of the other SM couplings, such as $\alpha_{weak}$, while also producing a small 
scaling violation, will not play too large of a numerical role here.) This implies that the ratio $R$ depends on $\sqrt s$ 
even for fixed values of $M/{\sqrt s}$ for any realistic NP signal process. In fact, as we will discuss below, we will 
find that it is always true that both these sources of scaling violation will force $R<1$ for any $\sqrt s >14$ TeV.  
This subsequently implies that even greater increases in integrated luminosities will be required to maintain the same 
relative mass reach than what is suggested by the naive scaling argument made above.  Specifically, we will show that, 
roughly speaking, over the $\sqrt s$ range of interest, the ratio $R$ scales as $\sim (\sqrt s)^{-p}$ with $p>0$, although 
this crude approximation breaks down over larger ranges of collision energy. Thus instead of the the naive scaling of 
the needed luminosity $L_{NS} \sim (\sqrt s)^2$, the actual luminosity scaling required to maintain the same relative mass 
reach is larger and is given approximately as $L \sim (\sqrt s)^{2+p}$.  As we will see below, for some processes $p$ can 
be as large as $\sim 0.6$ which implies a significant increase in the value of the desired luminosity.  Thus the value 
of $1/R$ tells us the {\it additional} multiplicative factor that one needs to apply, beyond that obtained from simple 
scaling, in order to maintain the same relative mass reach as we increase $\sqrt s$.  Of course in actuality we will 
find that $p$ is only approximately constant in each case and it, itself, also increases slowly with $\sqrt s$. However, 
we will see that the larger the roles of $\alpha_s$ and the gluon PDF are in the NP production cross section the greater 
the value of $p$ and the larger the requited integrated luminosity will be to maintain the scaled mass reach will be. 

Given these numerical results it will be the subject of further future studies to decide if the benefits of achieving 
this desired luminosity goal are worth the associated costs. Note that in the discussion below we will not address in 
any great detail the issue of the consequences to the various NP search reaches in the event that the corresponding 
integrated luminosity goals are not met at a future collider; we will, however, providing several examples. From these 
few cases it would seem that the impact of falling short of these goals, while limiting, still results in an enormous 
gain in search reach. However, more detailed work needs to be performed to fully access this very important problem. These 
are not new issues; discussions such as these took place over 30 years ago at the beginning of the SSC era{\cite {EHLQ}}.

\section{Effects of PDF Scaling Violations and $\alpha_s$ Running}

\begin{figure}[htbp]
\centerline{\includegraphics[width=5.0in,angle=90]{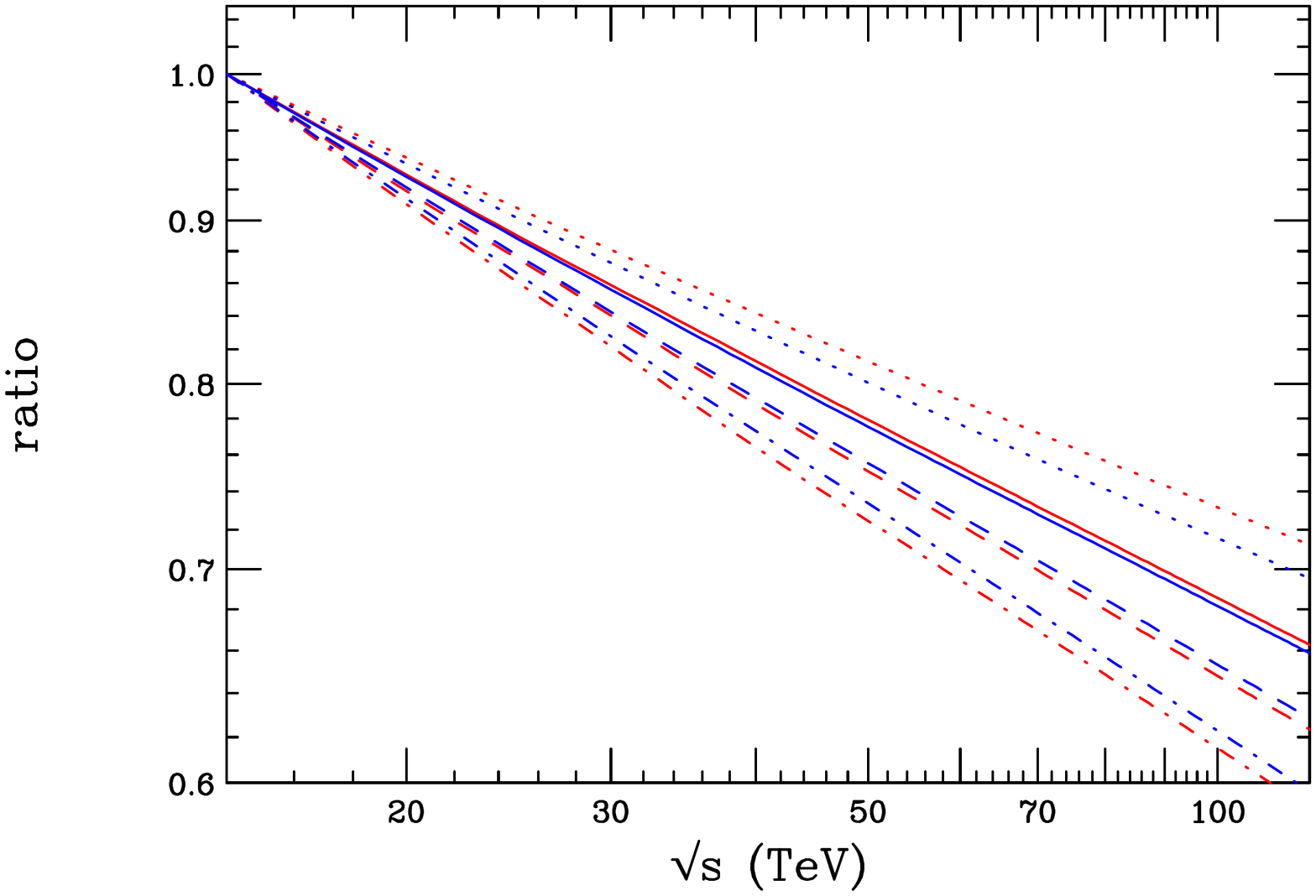}}
\vspace*{-3.0cm}
\centerline{\includegraphics[width=5.0in,angle=90]{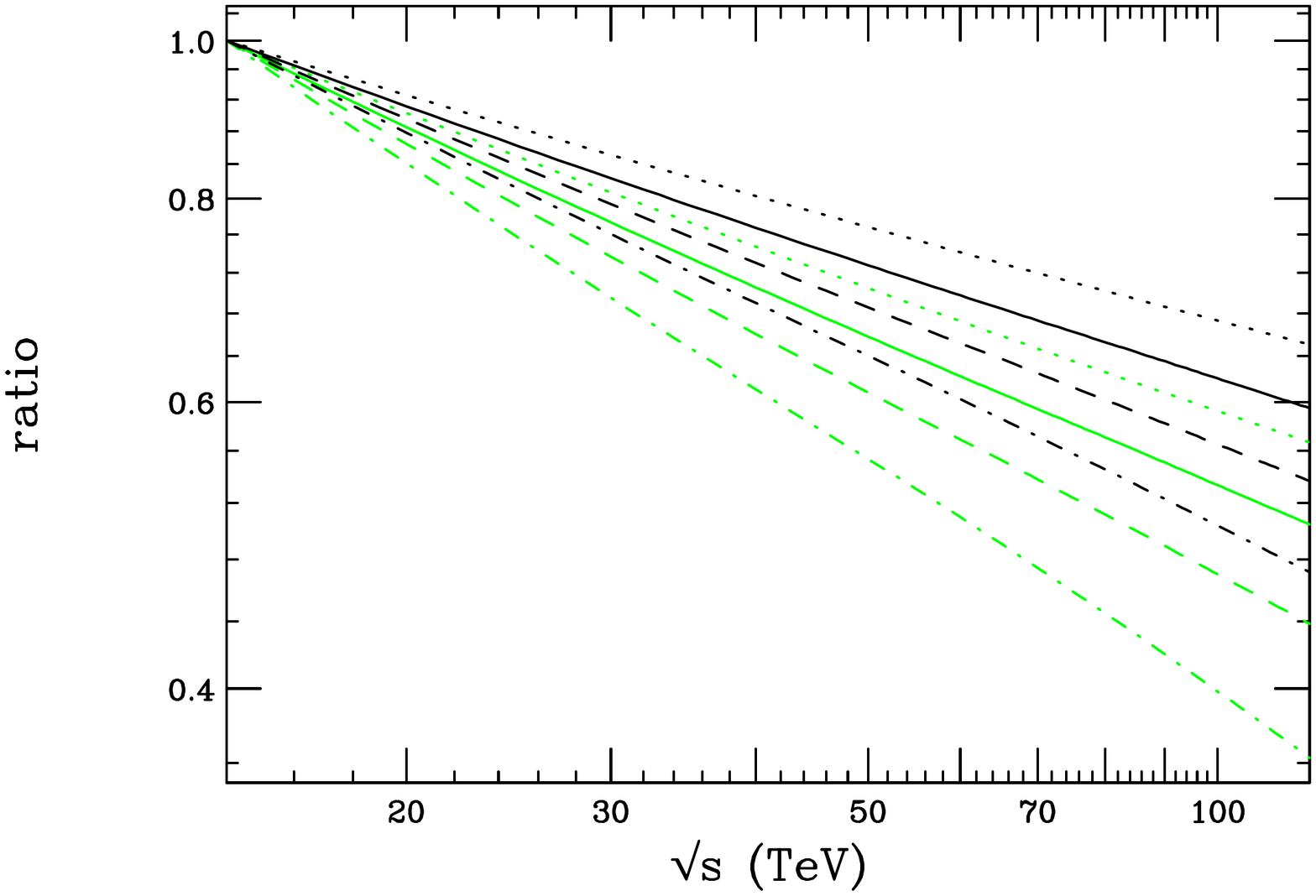}}
\vspace*{-1.90cm}
\caption{Ratios of the parton luminosities, defined in the text, as functions of $\sqrt s$ in comparison to those at 
$\sqrt s=14$ TeV. The dotted(solid, dashed, dash-dotted) curves in all case corresponds to fixed values of 
$M/\sqrt s=0.3(0.4,~0.5,~0.6)$. In the upper panel the red(blue) curves corresponds to the $U\bar U(D\bar D)$ cases, 
respectively, as discussed in the text. In the lower panel, the corresponding green(black) curves correspond to the 
cases $gg$ and $gQ$, as also discussed in the text, respectively.}
\label{fig1}
\end{figure}

To begin this discussion it is useful to examine how the ratio $R$ is influenced purely by the evolution of the PDFs 
themselves. In order to be definitive, we will employ the recently available default 5-flavor NNLO MMHT2014 set of 
PDFs{\cite {MMHT}} as well as their corresponding default NNNLO running value of $\alpha_s$ assuming that 
$\alpha_s(M_Z)=0.118$ in the numerical analysis that appears below. It should be noted, however, that the results we obtain 
can be easily shown to be quite insensitive to these particular choices. In order to access these effects,  we consider 
the standard integrated products of the parton densities, \ie, the parton luminosities:   
\begin{equation}
{\cal L}_{ij}= \int_\tau^1 ~dx ~\Big(f_i(x, M^2)f_j(\tau/x, M^2) + i\leftrightarrow j \Big)/x \,
\end{equation}
where $f_i$ is the relevant PDF, $\tau =M^2/s$ and with $M$ being identified with the partonic invariant mass at which 
the PDFs are evaluated. Depending upon the nature of NP, different combinations of the PDFs are usually involved. Here, 
to provide some specific examples we consider four sample cases: ($i$) we perform a sum over $i,j$ pairs corresponding to 
the product of $Q=2/3$ quarks and anti-quarks, denoted as $U\bar U = u\bar u+c\bar c$,  ($ii$) we instead sum over the 
corresponding products of $Q=-1/3$ quarks and anti-quarks, denoted as $D\bar D = d\bar d +s\bar s+b\bar b$, ($iii$) the 
product of the gluon density with a sum over all the quark and anti-quark densities, denoted as 
$qG = g(u+\bar u +d+\bar d +...)$, and lastly ($iv$) the product of two gluon densities, noted as $gg$. 
For each of these cases we fix the values of $M/\sqrt s$ and then determine the corresponding ratios 
${\cal L}(\sqrt s)/{\cal L}(\sqrt s=14 ~{\rm TeV})$ as $\sqrt s$ is increased. These ratios are, of course, all unity in 
the scaling limit.  The results of these calculations are shown in Fig.~\ref{fig1} to which we now turn.       

From the two panels in this Figure we learn several important things: ($a$) the ratio of luminosities in the range 
of interest scale roughly as $(\sqrt s)^{-q}$ with $q>0$ as expected from above which makes this dependence appear 
{\it almost}  linear in these log-log plots. {\footnote {Note $q$ is generally distinct from the parameter $p$ 
introduced above.}}  The particular value of $q$ depends upon the choice of the colliding partons 
as well as the specific choice of the value of $M/\sqrt s$. ($b$) In all cases we observe that as $M/\sqrt s$ increases 
so does the value of the `slope' $q$. ($c$) Since the gluon PDF evolves more quickly than do the valence or 
sea quarks, the slopes 
in the $gg$ case are seen to be greater than in the corresponding ones in the $gQ$ cases which are themselves greater 
than the corresponding ones observed in the $U\bar U$ and $D\bar D$ cases. In particular we see that in the $gg$ case the 
luminosity ratio for $M/\sqrt s=0.6$ falls off by a factor of $\sim 2.5$ as $\sqrt s$ increases to 100 TeV from 14 TeV 
due to the $Q^2$ evolution. This would mean that if {\it only} the PDFs mattered in the scale breaking then keeping the 
relative reach of $M/\sqrt s=0.6$ for NP fixed when going from 14 TeV to 100 TeV the luminosity would need to increase by 
a factor of $\sim 128$.  ($d$) The $U\bar U$ combination is also seen to evolve somewhat more quickly with $M/\sqrt s$ than 
does the corresponding $D\bar D$ PDF combination. Of course, there are {\it other} sources of scaling violation which can 
potentially change the value of 'partonic' slope $q$ to that of interest for a specific physical process, $p$.

\begin{figure}[htbp]
\centerline{\includegraphics[width=5.0in,angle=90]{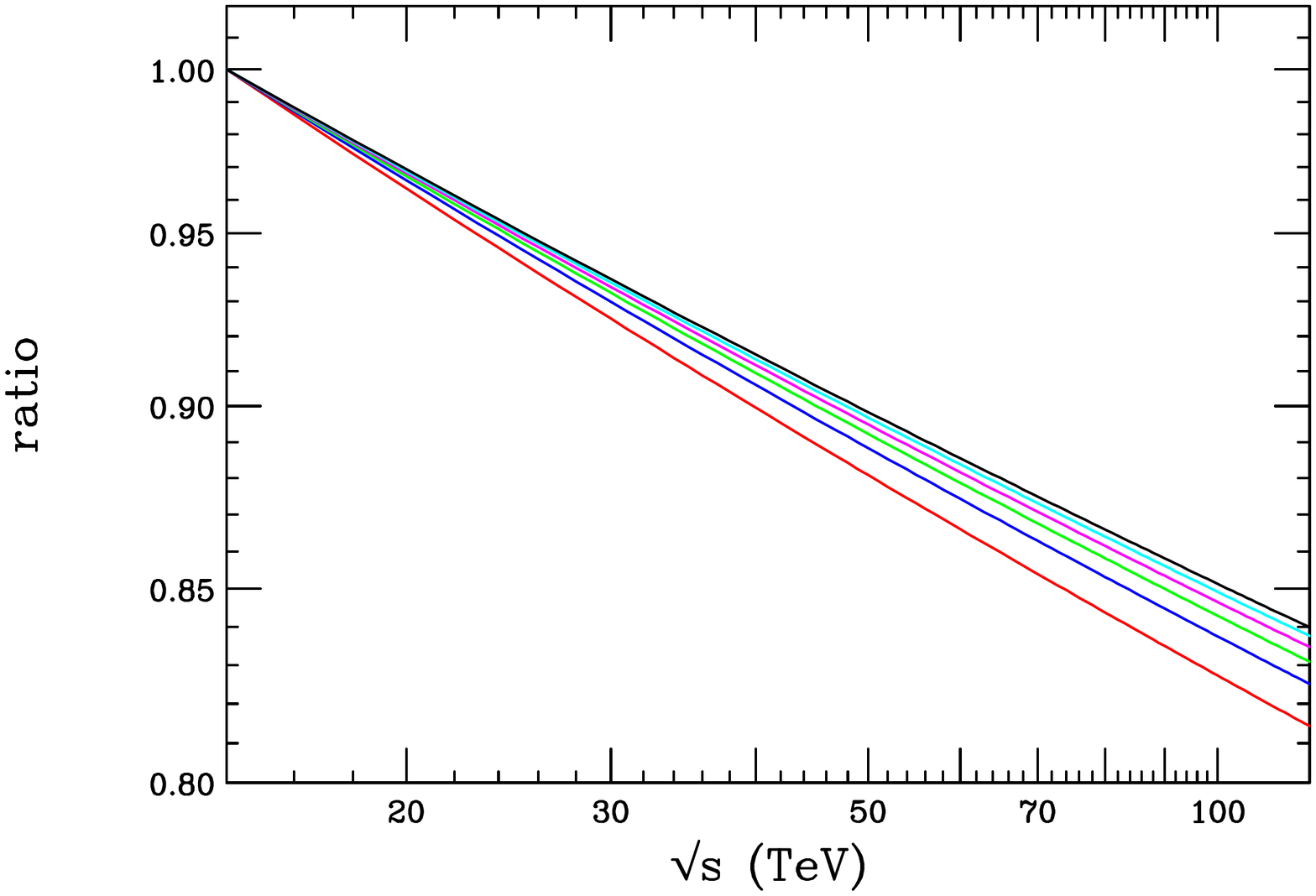}}
\vspace*{-1.90cm}
\caption{The ratio of the value of $\alpha_s(M)$, for fixed $M/\sqrt s$, as a function of $\sqrt s$ to that at 14 TeV. 
From bottom to top the curves correspond to $M/\sqrt s=0.1$ to 0.6 in steps of 0.1.}
\label{fig0}
\end{figure}

In a similar manner one can examine the magnitude of the effect of scaling violations from the running of $\alpha_s(M)$, 
for fixed values of $M/\sqrt s$, by comparing its evolution as $\sqrt s$ increases beyond 14 TeV; Fig.~\ref{fig0} shows 
this result. Here we see that this evolution is (very) roughly constant for reasonable, yet fixed, values of $M/\sqrt s$. 
However, the slope is slightly less steep for larger $M/\sqrt s$ values and some non-trivial curvature away from a 
straight line behavior is clearly visible. How much the running of $\alpha_s$ will contribute to the overall scaling 
violation in a given process strongly depends upon the power in which it appears in the relevant subprocess cross 
sections, $\hat \sigma \sim \alpha_s^m$. In the cases we will examine below we have $m=$0, 1 or 2.  

\section{Sample Benchmark New Physics Scenarios}

We now turn to an examination of how the production cross sections and mass reach associated with different specific 
benchmarks of NP will scale as we increases $\sqrt s$ beyond 14 TeV along the lines discussed above by employing the ratio 
$R$. This survey is not meant to be in any way exhaustive but only to be indicative of the possible range of the values 
of the ratio $R$ and to demonstrate what might be expected from going beyond the naive scaling arguments. Based on the 
discussion above it is clear that the results we obtain will depend upon which subset of the PDFs are dominant and the 
role that $\alpha_s$ plays in the relevant NP production process, generally at LO, as NLO and higher order corrections 
will be sub-leading.

\begin{figure}[htbp]
\centerline{\includegraphics[width=5.0in,angle=90]{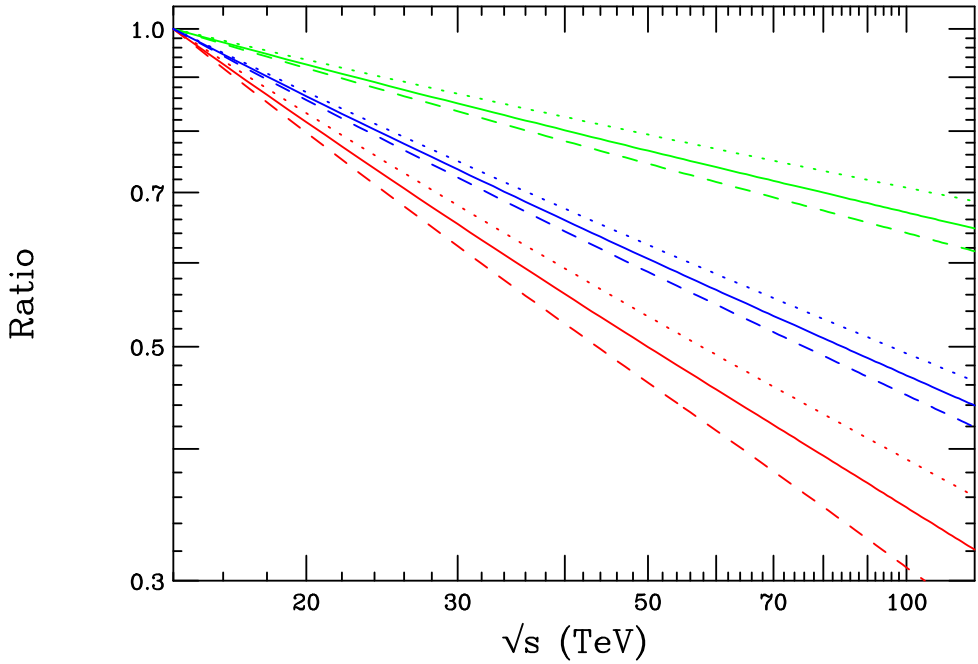}}
\vspace*{-1.90cm}
\caption{The ratio $R$ of scaled cross sections defined in the text as functions of $\sqrt s$. Here 
the dotted(solid, dashed) green curves corresponds to values of $M/\sqrt s=0.3(0.4,~0.5)$ for inclusive 
$W'^{\pm}$ production followed by leptonic decay with $M=M_{W'}$. The blue(red) curves are for inclusive heavy 
quark pair-production via $q\bar q(gg) \to Q\bar Q$; in this case $M$ is identified with the heavy quark mass 
and the dotted(solid, dashed) curves corresponds to values of $M/\sqrt s=0.15(0.20,~0.25)$, respectively.}
\label{fig2}
\end{figure}

We begin by considering the Drell-Yan production of the $W'^\pm$ gauge boson in the Sequential SM{\cite {rev}} in 
the narrow width approximation (NWA) followed by its subsequent leptonic decay. In this case the $\alpha_s$ corrections 
to both the cross section and leptonic branching fractions appear only at NLO (but are included here) and so the 
scaling behavior is expected to be dominated by that associated with the $U$- and $D$-type PDFs. Fig.~\ref{fig2} shows 
the scaled cross section ratio for this process assuming that $M_{W'}/\sqrt s=0.3,~0.4$ or 0.5. These values were chosen 
since the expected the reach for the SSM $W'$ at the 14 TeV LHC typically lies in the 5-7 TeV range. Indeed, the 
behavior of the ratio $R$ in this case is observed to generally follow that produced by a weighted combination 
of that for the $U\bar U$ and $D\bar D$ PDFs as we see by comparison with the top panel in Fig.~\ref{fig1}.  This means, 
\eg, that the required luminosity to maintain the same value of the scaled search reach is only modestly larger than 
suggested by the scaling estimate.  Note that as expected $R$ behaves in all cases roughly as $(\sqrt s)^{-p}$, \ie, 
is almost linearly on this log-log plot.  Table~\ref{rates} shows that the required integrated luminosity scaling for a 
$\sqrt s=$100 TeV collider in comparison to the 14 TeV LHC to be in the $\sim 74-80$ range for this process which 
is {\it only} $\sim 50\%$ larger that than implied by the scaling limit.      
 
Interestingly, if we consider the $\sqrt s=100$ TeV collider as a specific example, it is easy to access the cost in 
the search reach for the $W'$ if the `required' integrated luminosity goal is not reached. We find{\cite {me}} that 
for integrated luminosities in the 1-1000 ab$^{-1}$ range the discovery reach increases/decreases by $\sim 7.60$ TeV 
for every factor of 10 of integrated luminosity gained or lost. The search reach is found to be $\sim 39.1$ TeV assuming 
10 ab$^{-1}$ is available so this is roughly a factor of $\sim 20\%$ for an order of magnitude change in luminosity.

A second interesting example is provided by new heavy quark ($Q$) pair production which proceeds by $gg,q\bar q$ fusion 
at LO. Although the actual cross section for $Q\bar Q$ production is a PDF-weighted combination of these two contributions,  
it is interesting to consider both of these processes separately since they depend on different PDFs yet both occur 
at $O(\alpha_s^2)$ in LO. Fig.~\ref{fig2} shows the $R$ ratio for both of these processes assuming that 
$M(=M_Q)/\sqrt s=0.15-0.25$ corresponding to $M_Q \sim 2-3$ TeV at the 14 TeV LHC. Here we see that $R$ falls much faster 
than do the corresponding relevant PDFs in either case due to the additional scaling violation associated with the running of 
the $\alpha_s$ coupling. Table~\ref{rates} shows that in these cases the required integrated luminosity scaling for a 
$\sqrt s=$100 TeV collider to be in the $\sim 103-115$ range for the $q\bar q$-initiated process and $\sim 130-165$ for the 
$gg$-initiated process, which is expected to be dominant at large values of $\sqrt s$.  If we roughly require 100 signal 
events at a $\sqrt s=100$ TeV collider as a benchmark signal rate for purposes of comparison (note not for a discovery or 
to set a limit) this corresponds to an heavy quark mass of $\sim 9.42(12.40,~15,57,~18.82)$ TeV for integrated luminosities 
of 1(10, 100, 1000) ab$^{-1}$, respectively,  which displays how the mass reach scales with luminosity variations in 
the interesting range. We see that this corresponds to roughly a factor of 
$\sim 20\%$ change in mass reach for an order of magnitude change in the integrated luminosity.  

\begin{figure}[htbp]
\centerline{\includegraphics[width=5.0in,angle=90]{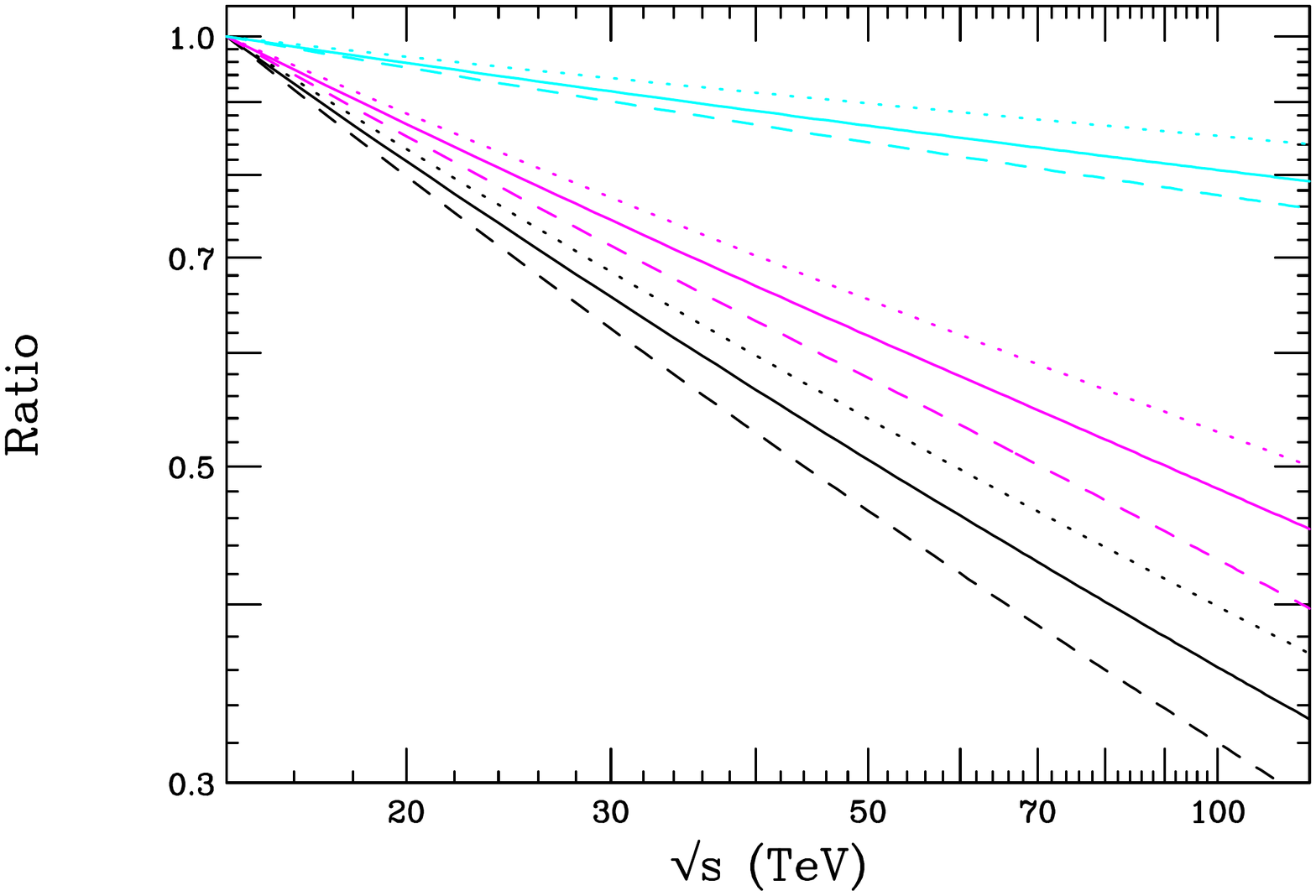}}
\vspace*{-1.90cm}
\caption{The ratio $R$ of scaled cross sections defined in the text as functions of $\sqrt s$ as in the previous 
figure. The cyan curves are for the production of heavy vector-like, isodoublet lepton pairs, $q\bar q \to L^+L^-$ 
where $M=M_L$ and the dotted(solid, dashed) curves correspond to values of $M/\sqrt s=0.03(0.05,~0.07)$, respectively. 
The red curves are for the resonant single production of a color-triple excited quark, with $M=M^*$, in $gq$ fusion 
assuming that $M/\sqrt s=0.4(0.5,~0.6)$ as discussed in the text. The corresponding black curves are for scalar, 
color-triplet leptoquark pair production summed over both the $gg$ and $q\bar q$ channels taking $M=M_{LQ}$ and 
assuming $M/\sqrt s=0.15(0.20,~0.25)$, respectively.}
\label{fig3}
\end{figure}

Several more NP processes are considered in Fig.~\ref{fig3} the first being heavy vector-like lepton ($L$) pair 
production via $q\bar q$ annihilation through $\gamma,Z$ exchange, ignoring the potential contribution of the 
$gg$-fusion, loop-induced process{\cite {lepton}} since it is more model dependent. To be specific, we will consider 
the case of a singly-charged, weak isodoublet lepton, as might appear in an $E_6$-type framework{\cite {it}} although 
the results we obtain are quite independent of this particular 
choice. Since this production cross section in this case is rather small and searches for such heavy leptons are 
dominated by large SM backgrounds from inclusive $W^+W^-$ production, we restrict ourselves to rather low leptonic 
mass values, \ie, $M(=M_L)/\sqrt s=0.03-0.07$, corresponding to heavy lepton masses well below 1 TeV at LHC14. Since 
these $M/\sqrt s$ values are so low and $\alpha_s$ enters only at NLO for this process we expect the $\sqrt s$ dependence 
of $R$ as well as its deviations from unity to be rather weak. This is exactly what we find by looking at Fig.~\ref{fig3}.  
This means, \eg, that the required luminosity to maintain the same value of the scaled search reach is only a bit 
larger than suggested by the scaling estimate in this scenario.  Table~\ref{rates} shows that the required integrated 
luminosity scaling for a $\sqrt s=$100 TeV collider to be in the $\sim 60-66$ range for such heavy leptons.   For a 
$\sqrt s=100$ TeV collider, and taking 100 events as a standard for comparison of the potential mass reach,  we find 
this event rate corresponds to heavy lepton masses of  $\sim 400(625, ~904, ~1218)$ GeV assuming luminosities of 
1(10, 100, 1000) ab$^{-1}$ so that variations of a factor of 10 produce a roughly $\sim 30\%$ change in the approximate 
mass reach, a value not too dissimilar from the previous case.         

\begin{figure}[htbp]
\centerline{\includegraphics[width=5.0in,angle=90]{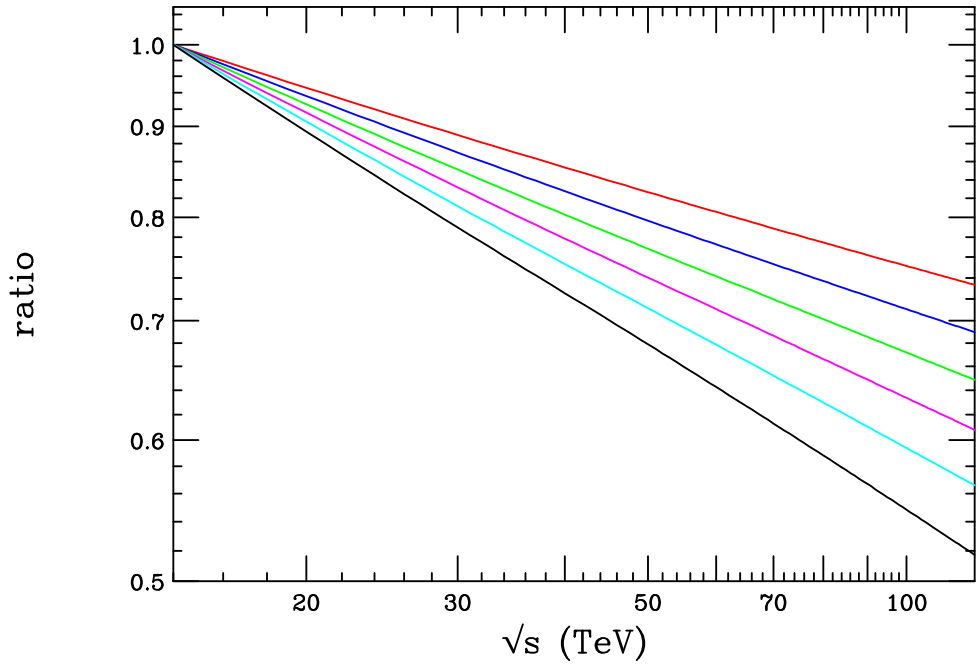}}
\vspace*{-3.0cm}
\centerline{\includegraphics[width=5.0in,angle=90]{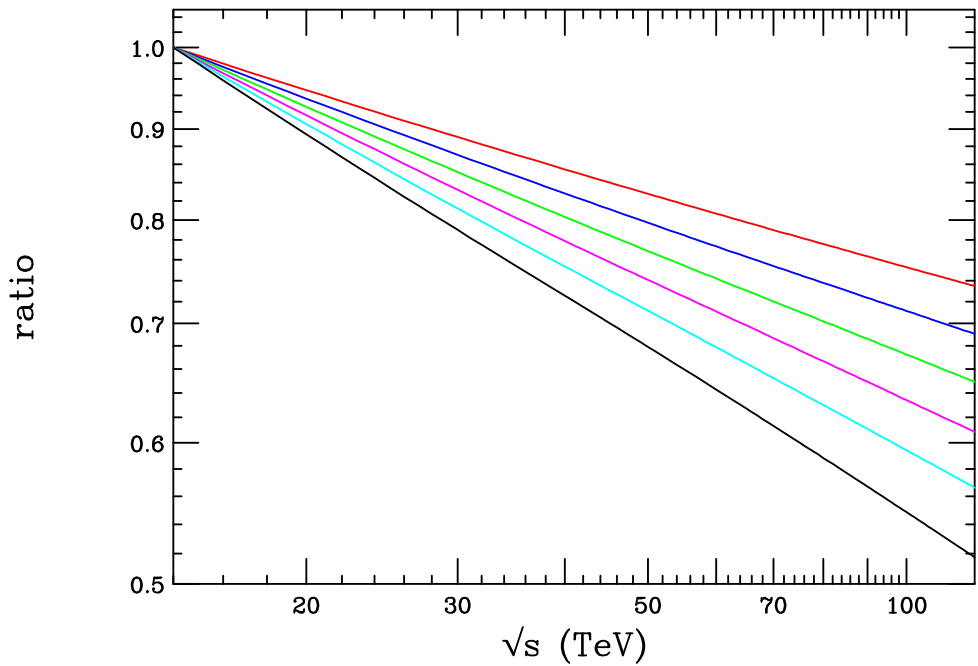}}
\vspace*{-1.90cm}
\caption{The ratio $R$ for TeV-scale black hole production assuming $n=2$ (top) or 6 (bottom) additional flat dimensions 
with, from top to bottom $M_{P}/\sqrt s=0.2-0.7$ in steps of 0.1, respectively.}
\label{fig4}
\end{figure}

Another example provided by Figure~\ref{fig3} is that of scalar (spin-0), color triplet leptoquark (LQ) {\cite {LQ}} 
pair production which, like heavy quark production arises from both $q\bar q$- and $gg$-initiated processes which 
we combine into the total cross section in the presentation below. Since in LO these are also $\alpha_s^2$ processes 
but differ in kinematic detail from the corresponding $Q\bar Q$ ones (due to their spin-0 nature) we expect results 
which are similar to but quantitatively different from those found for heavy quark production.  This is exactly what 
we observe in Fig.~\ref{fig3}. Table~\ref{rates} shows that in this case the required integrated luminosity scaling 
for a $\sqrt s=$100 TeV collider to be in the $\sim 127-160$ range for such heavy leptoquarks, a value not very different 
than that for $gg$-initiated heavy quark production. For a $\sqrt s=100$ TeV collider, and taking 100 events as a minimal 
search criterion as above then the mass reach is found to be $\sim 6.60(8.97, ~11.70, ~14.70)$ TeV assuming integrated 
luminosities of 1(10, 100, 1000) ab$^{-1}$ so that variations by a factor of 10 in integrated luminosity 
again produce a roughly $\sim 30\%$ change in the mass reach.

A last example shown in Fig~\ref{fig3} is that of single, resonant, excited quark ($q^*$) (with mass $M_{q^*}$) 
production in $gq(\bar q)$-fusion{\cite {excited}} which proceeds via an effective dimensional-5 operator with a cross 
section that is proportional to $\alpha_s$. Here we also require that the each of the jets from the decay of the $q^*$ 
satisfy $\eta_j<0.5$ to reduce the QCD backgrounds. 
Explicitly, in the NWA one finds that in the case the partonic cross section 
behaves like $\hat \sigma ~\sim (\alpha_s/\Lambda^2)~(M_{q^*}^2/s)$ times a product of the appropriate PDFs where 
$\Lambda$ is identified with the `compositeness' scale associated with the dim-5 operator. 
Provided we keep both $M_{q^*}^2/s$ and $\Lambda^2/s$ fixed as $\sqrt s$ increases, then in the scaling limit the 
ratio $R$ is again always unity. In the real world, since the cross section only involves a single power of the 
strong coupling and the relevant PDFs are of the $gQ$ type we expect $R$ to have an intermediate behavior between 
that observed for heavy quark production and that found for $W'$ production and that is indeed the situation revealed 
in Fig.~\ref{fig3}. 
Here we assume that $M_{q^*}/\sqrt s=0.4-0.6$ given the present 8 TeV and the anticipated 14 TeV LHC search reaches 
for excited quarks. Table~\ref{rates} shows that in this case the required integrated luminosity scaling for a 
$\sqrt s=$100 TeV collider to be in the $\sim 96-116$ range for the heavy excited quark scenario. Interestingly, if 
we require 100 signal events at a $\sqrt s=100$ TeV collider as a benchmark signal rate as well as $\Lambda=M_{Q^*}$ 
for simplicity, this corresponds to an excited quark mass of 36.1(44.7, 52.8, 60.3) TeV for integrated luminosities 
of 1(10, 100, 1000) ab$^{-1}$ which shows how the reach scales with luminosity in the region of interest. This roughly 
corresponds to a factor of $\sim 20\%$ for an order of magnitude change in the integrated luminosity.

\begin{table}
\centering
\begin{tabular}{|l|c|c|} \hline\hline
Particle    &$M/\sqrt s$ &L(100)/L(14)\\
\hline

$W'$   & 0.30   &    74.3  \\
$W'$   & 0.50   &    79.6  \\
Q($q\bar q$) & 0.15  & 103.6   \\
Q($q\bar q$) & 0.25  & 113.5   \\
Q(gg)  & 0.15   &   130.5  \\
Q(gg)  & 0.25   &   165.2    \\
L      & 0.03   &    59.9   \\
L      & 0.07   &    65.9   \\
LQ     & 0.15   &   127.9   \\
LQ     & 0.25   &   159.4   \\
$q^*$  & 0.40   &    96.5  \\ 
$q^*$  & 0.60   &   118.8  \\
BH6    & 0.20   &    67.8   \\
BH6    & 0.70   &    93.0   \\
\hline\hline
\end{tabular}
\caption{Ratios of the integrated luminosity required at a 100 TeV FHC to maintain the same relative 
mass reach, $M/\sqrt s$,  as that at the 14 TeV LHC for various types of NP as discussed in the text.}
\label{rates}
\end{table}

A final example of NP that we consider is the production of TeV-scale black holes (BH) of mass $M_{BH}$ that arise in 
theories with extra dimension{\cite {BH}}. In the simplest toy models with additional flat dimensions, BH production has 
a simple threshold at the $n+4$-dimensional Planck mass, $M_P$, with a continuum above this value and has a partonic 
cross section that scales as $\hat \sigma_{BH} \sim (M_{BH}/M_P)^{{2}/{1+n}}~M_P^{-2}$. It is also usually assumed that 
the products of {\it all} the PDFs enter here since the collision of any two partons can make a BH with the same 
efficiency given sufficient collision energy.  This implies that the total BH cross section then behaves as
\begin{equation}
\sigma_{BH} \sim \int_{\tau_{min}}^1 ~d\tau ~\int_\tau^1 ~dx ~\Pi_{ij} ~\Big(f_i(x, M^2)f_j(\tau/x, M^2)/x\Big)~\hat \sigma_{BH}  \, 
\end{equation}
where here $\tau_{min}=M_P^2/s$ and so the relevant scaled mass reach to consider in this case is just the quantity 
$M_P/\sqrt s$. the factor $\Pi_{ij}$ represents the sum over all pair-wise products of the PDFs, $f_i$, as employed 
in the previous section. The BH mass is then identified with the resulting partonic invariant mass above the value of 
$M_P$. Since the cross section for BH production can in principle be very large but may also experience substantial 
suppression factors of various kinds we here consider the wide range of values for $M_P/\sqrt s=0.2-0.7$ and the 
specific cases of $n=2,6$. The results we obtain are only very weakly dependent on the specific value of $n$ as we see 
from the curves shown in Fig.~\ref{fig4} but they are dependent on the specific chosen value of $M_P/\sqrt s$. Since 
the pair-wise product of all the PDFs enter here (and there are no $\alpha_s$ factors present) we might roughly expect 
the behavior of the $R$ ratio in this case to be similar that of the valence PDFs $U\bar U$ and $D\bar D$ and this is 
essentially what we see here.  Table~\ref{rates} shows that in this case the required integrated luminosity scaling for 
a $\sqrt s=$100 TeV collider for an $n=6$ BH to be in the $\sim 67-93$ range.  Again, for a $\sqrt s=100$ TeV collider, 
and taking as usual 100 events as a minimal search criterion then the mass reach for $n=6$ is found to be 
$\sim 47.3(54.7,~ 61.6, ~67.9)$ TeV assuming luminosities of 1(10, 100, 1000) ab$^{-1}$ so that variations of a factor 
of 10 in integrated luminosity would produce a roughly $\sim 15-20\%$ change in the BH mass reach.

\section{Luminosity Impact on Reaches}

\begin{figure}[htbp]
\centerline{\includegraphics[width=5.0in,angle=90]{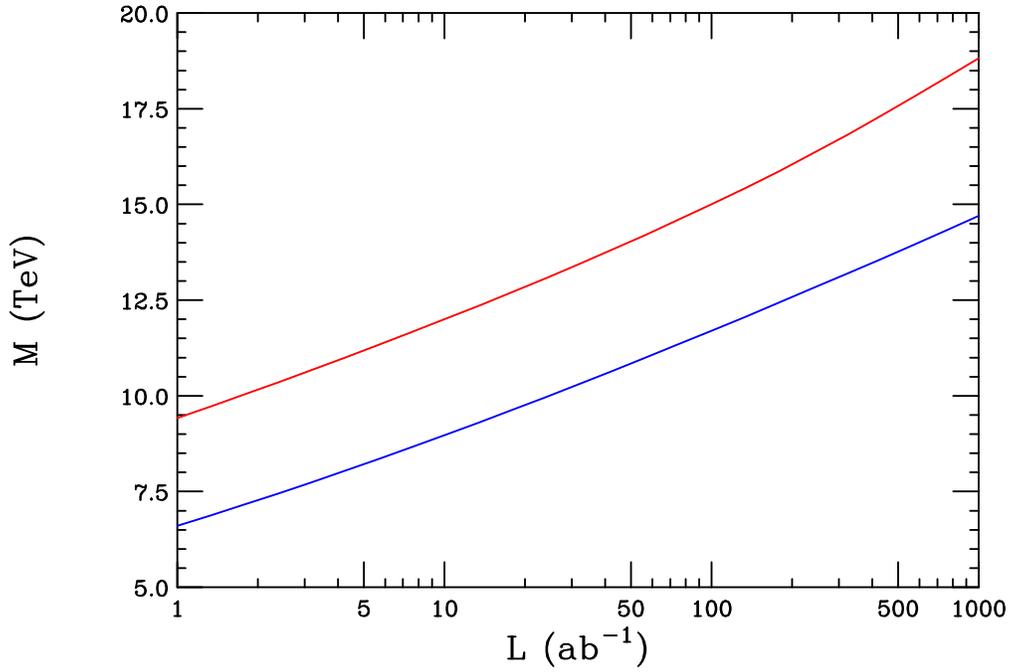}}
\vspace*{-3.0cm}
\centerline{\includegraphics[width=5.0in,angle=90]{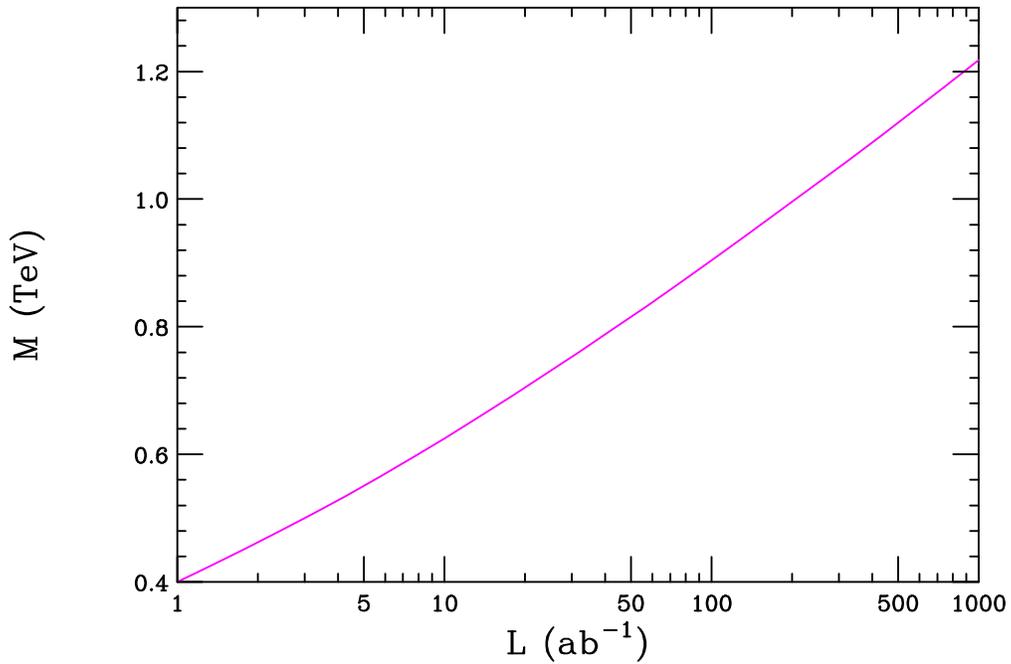}}
\vspace*{-1.90cm}
\caption{Luminosity dependence of the various mass reaches discussed in the previous section for a $\sqrt s=100$ TeV 
collider. The top panel shows this dependence for scalar leptoquarks (blue) and heavy quarks (red) while the lower 
panel shows that for vector-like leptons produced in $q\bar q$ collisions.}
\label{fig5}
\end{figure}

\begin{figure}[htbp]
\centerline{\includegraphics[width=5.0in,angle=90]{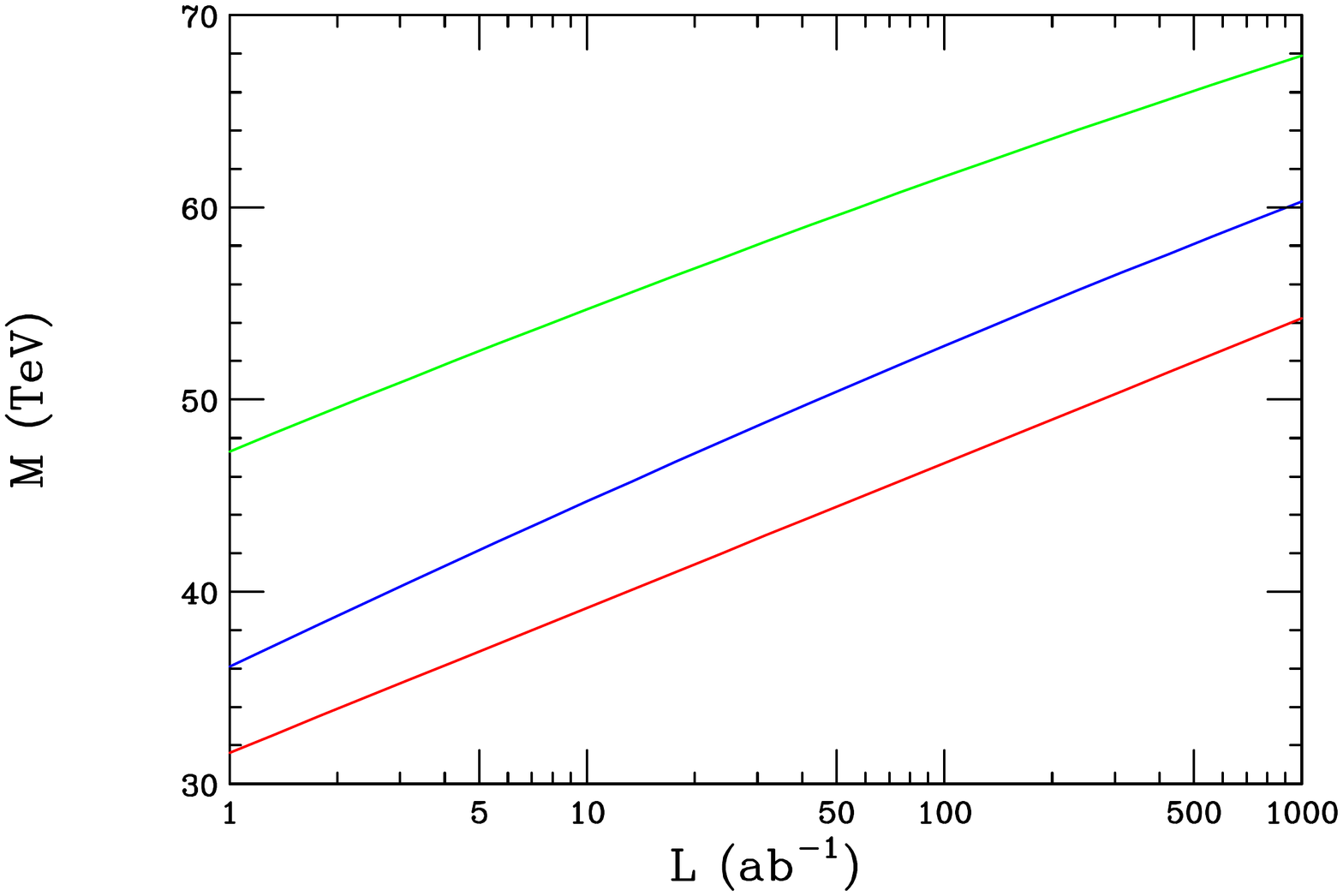}}
\vspace*{-1.90cm}
\caption{Same as the previous Figure, but now, from top to bottom, for a $n=6$ BH, an exited quark and for a SSM $W'$.}
\label{fig6}
\end{figure}

In this Section we will briefly examine the impact of achieving luminosities different from those required to maintain the 
mass reach scaling law discussed above; to be specific we consider the case of a $\sqrt s=100$ TeV collider. This issue is most 
easily analyzed by considering the mass reach results shown in Figs.~\ref{fig5} and ~\ref{fig6} which employ the same physics 
examples discussed above. Specifically, these Figures show how the mass reaches for the six new physics scenarios considered in 
the previous section will scale with the integrated luminosity achieved. As alluded to above we see that in all cases the mass 
reach is found to depend almost linearly, roughly speaking, with the log of the luminosity in the range of interest. The important 
point demonstrated by these two Figures is that even if there is {\it no} luminosity gain whatsoever over the LHC, the mass reach 
of a $\sqrt s=100$ TeV collider is still quite good.

To make an even more direct comparison, we note in Table~\ref{rates} the mass reaches for these six NP scenarios for a 
$\sqrt s=100$ TeV collider assuming integrated luminosities of 1 ab$^{-1}$, comparable to the 14 TeV LHC and so no gain in 
luminosity, and for 100 ab$^{-1}$, which is very roughly the average value of the luminosity required to maintain mass reach 
scaling as are given in Table~\ref{rates}. In a sense, this might be considered to be the worst case scenario. Here, we see 
explicitly that the mass reach reduction experienced in this rather extreme situation, due to a factor of 100 times less 
integrated luminosity, is in all but one scenario (where the factor of 100 employed is too large) is less than 30-40$\%$.  
Thus we conclude that while not reaching the desired value of the integrated luminosity given by the scaling requirement 
does result in a reduction in mass reach, it is found to be not a very serious loss for the NP scenarios considered here. Of 
course more detailed work needs to be done to more fully understand the impact of a reduced integrated luminosity on the 
various NP searches.

\begin{table}
\centering
\begin{tabular}{|l|c|c|} \hline\hline
Particle    & 1 ab$^{-1}$ & 100 ab$^{-1}$\\
\hline

$W'$   & 31.6  & 46.7    \\
Q      & 9.42 &  15.57   \\
L      & 0.40 &  0.90   \\
LQ     & 6.60 &  11.70  \\
$q^*$  & 36.1  & 46.7   \\ 
BH6    & 47.3  & 67.9    \\
\hline\hline
\end{tabular}
\caption{Comparison of the mass reaches in TeV for various NP scenarios at a $\sqrt s=100$ TeV collider assuming an integrated 
luminosity of 1 ab$^{-1}$ and 100 ab$^{-1}$, very roughly the average value required for mass reach scaling.}
\label{rates2}
\end{table}

\section{Discussion and Conclusions}

Proposed future hadron colliders will have a vastly improved mass reach for new physics in comparison to that of the 
14 TeV LHC. In this paper we have examined how the required integrated luminosity for such machines must scale in 
order to maintain the same {\it relative} mass reach, $M/\sqrt s$, as at the LHC for various kinds of new physics benchmark 
models as $\sqrt s$ increases beyond 14 TeV. For a $\sqrt s=100$ TeV collider, the scaling limit requires the integrated 
luminosity to be $\sim 50$ times larger than at the 14 TeV LHC for {\it all} types of NP and for {\it all} fixed values 
of $M/\sqrt s$. In contrast to this, due to inherent scaling violations of the PDFs and, in some cases that of the strong 
 coupling, $\alpha_s$, we obtain a rather wide range of possible values, $\sim 60-165$, for various kinds of NP benchmarks 
and for different assumed values of $M/\sqrt s$. All of these integrated luminosity values are larger, and in some cases  
significantly larger, than that given by the scaling argument. The costs of reaching these types of integrated luminosity 
goals will be quite high and it will be up to further detailed studies to decide how well they can or should be met at 
any future hadron collider. Here we have made some crude estimates that indicate that in all the NP scenarios examined a  
reduction in the integrated luminosity by one (two) order(s) of magnitude around the relevant NP mass ranges of interest would 
result in search reach degradation of, roughly, only $\sim 15-30(30-40)\%$ for a $\sqrt s=100$ TeV collider. However, even if 
only this lower luminosity were to be achieved the gain in mass reach obtained by going to larger values of $\sqrt s$ remains 
very substantial.

\section*{Acknowledgments}

The author would like to thank both M. Breidenbach and particularly B. Richter for continually emphasizing to him 
the importance of the luminosity scaling issue for future hadron colliders and pushing this project along. He would 
also like to thank J.L. Hewett for many discussions related to the NP choices made in this analysis as well as Ken Lane 
for more general discussions. This work was supported by the Department of Energy, Contract DE-AC02-76SF00515.

\end{document}